\documentclass[twocolumn,showpacs,amsmath,amssymb,prl]{revtex4}

\usepackage{graphicx}   
\usepackage{dcolumn}    
\usepackage{bm}         

\newcommand{\abs}[1]{\left\vert#1\right\vert}   
\newcommand{\mr}{\mathrm}                       

\begin{document}
\title{Critical voltage of a mesoscopic superconductor}


\author{R. S. Keizer$^{1,\dagger}$, M. G. Flokstra$^{2,\dagger}$, J. Aarts$^2$ and T. M. Klapwijk$^1$}
\affiliation{$^\dagger$these authors contributed equally to the contents of this paper\\
$^1$Kavli Institute of NanoScience, Delft University
of Technology, Lorentzweg 1, 2628 CJ Delft, The Netherlands \\
$^2$Kamerlingh Onnes Laboratory, Universiteit Leiden, 2300 RA
Leiden, The Netherlands}


\date{\today}

\begin{abstract}
We study the role of the quasiparticle distribution function $f$
on the properties of a superconducting nanowire. We employ a
numerical calculation based upon the Usadel equation. Going beyond
linear response, we find a non-thermal distribution for $f$ caused
by an applied bias voltage. We demonstrate that the even part of
$f$ (the energy mode $f_L$) drives a first order transition from
the superconducting state to the normal state irrespective of the
current.
\end{abstract}

 \maketitle

\begin{figure}[b]
  \includegraphics[width=8cm]{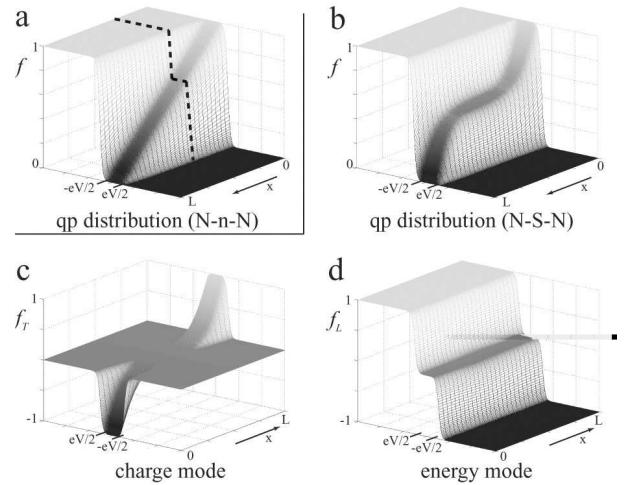}\\
  \caption{Quasiparticle distribution function $f(x,\varepsilon)$ as function of energy $\varepsilon$ and
   position $x$ for a normal wire (a) and a superconducting wire
  (b) between normal metallic reservoirs for $k_BT\ll
  eV<\Delta_0$, with (c) and (d) the decomposition of (b) into the charge mode $f_T$ and energy mode $f_L$.}
  \label{fig:Distributions}
\end{figure}

The energy distribution function of quasiparticles in a normal
metal is under equilibrium conditions given by the Fermi-Dirac
distribution $f_0$. In recent years it has been demonstrated that
in a voltage ($V$)-biased mesoscopic wire (length $L$) a two-step
non-equilibrium distribution develops \cite{Pothier1997} with
additional rounding by quasiparticle scattering due to spin-flip
and/or Coulomb interactions \cite{Birge2003}. Figure
\ref{fig:Distributions}a shows the distribution, which resembles
two shifted Fermi-Dirac functions:
\begin{equation}
  f(x,\varepsilon)=(1-x)f_0(\varepsilon+eV/2)+xf_0(\varepsilon-eV/2)
\end{equation}
with $\varepsilon$ the quasiparticle energy and $x$ the coordinate
along the wire. For strong enough relaxation ($L \gg L_\phi$, with
$L_\phi$ is the phase coherence length) and/or high temperatures
($k_BT \gg eV$) the distribution returns to a Fermi-Dirac
distribution with a local effective temperature.

If the normal wire is replaced by a superconducting wire, the
attractive interaction between electrons leads to the
superconducting state. The questions we address here are how the
distribution function is modified (for a typical result see Fig.
\ref{fig:Distributions}b) and how this affects observable
properties such as the current-voltage characteristics of the
system and the breakdown of the superconducting state. To relate
the distribution function to observable quantities, it is
convenient to separate the symmetric part $f_L$ (energy mode) from
the asymmetric part $f_T$ (charge mode) which each have a
different spatial and spectral form (Fig. \ref{fig:Distributions}c
and d). In particular we will show that the breakdown is
characterized by a voltage rather than by a current; in other
words, the system cannot be trivially treated as two resistors
modelling the normal- to supercurrent conversion, with a
superconducting element characterized by its depairing current
in-between.

The transport and spectral properties of dirty superconducting
systems ($\ell_e\ll\xi_0$, with $\ell_e$ the elastic mean free
path and $\xi_0$ the superconducting phase coherence length) are
described by the quasiclassical Green functions obeying the Usadel
equation \cite{Kopnin2001}. For out of equilibrium systems we use
the Keldysh technique in Nambu (particle-hole) space. We look at
s-wave superconductors (singlet pairing) without any
spin-dependent interactions. The Usadel equation then takes the
form $\hbar D \nabla(\check{G}\nabla\check{G}) =
-i[\check{H},\check{G}]$, where the check notation ($\check{G}$)
denotes a $4 \times 4$ matrix, $D$ is the diffusion constant,
$\nabla$ is the spatial derivative \cite{Gauge} and we neglect any
inelastic process. The elements of $\check{G}$ and $\check{H}$,
when split up in Keldysh space, are $2 \times 2$ matrices in Nambu
space, denoted by a hat:
\begin{equation}\label{eqn:Keldyshmat}
  \check{G} = \left( \begin{array}{cc} \hat{G}^R & \hat{G}^K \\
              0 & \hat{G}^A \end{array} \right) \mbox{, }
  \check{H} = \left( \begin{array}{cc} \hat{H} & 0 \\
              0 & \hat{H} \end{array} \right)
\end{equation}
Here, $\hat{G}^R$ and $\hat{G}^A$ are the retarded and advanced
components describing equilibrium properties and $\hat{G}^K$ is
the Keldysh component which describes the non-equilibrium
properties. Their elements are the quasiclassical
(energy-dependent) normal and anomalous Green functions and, for
the Keldysh component only, the quasiparticle distribution
functions (which take account of the non-equilibrium). For the
Hamiltonian $\hat{H}$ we write:
\begin{equation}
  \hat{H} = \left( \begin{array}{cc} \varepsilon & -\Delta \\
                   \Delta^* & -\varepsilon \end{array} \right) \\
\end{equation}
\noindent where $\varepsilon$ is the (eigen)energy and the chosen
gauge is such that the pair potential $\Delta$ is in equilibrium a
real quantity, $\Delta=\Delta^*$.
The matrix Green function $\check{G}$ satisfies the normalization
condition $\check{G}\check{G} = \check{1}$, leading to
$\hat{G}^R\hat{G}^R = \hat{G}^A\hat{G}^A = \hat{1}$ and
$\hat{G}^R\hat{G}^K + \hat{G}^K\hat{G}^A = \hat{0}$. If
superconducting reservoirs in the system are kept at zero voltage
(avoiding AC Josephson effects), $\hat{G}^K$ can be written as
$\hat{G}^K = \hat{G}^R \hat{f} - \hat{f} \hat{G}^A$. Here
$\hat{f}$ is the diagonal generalized distribution number matrix
of the quasiparticles in Nambu space. To relate $\hat{f}$ to
observable quantities we decompose it into an even part (or
energy/longitudinal mode) and an odd part (or charge/transverse
mode) in particle-hole space: $\hat{f} = f_L \tau_0 + f_T \tau_3$,
where $\tau_i$ are the Pauli matrices in particle-hole
space \cite{Schmid197A}. 
The full distribution function is retained by: $2f(x,\varepsilon)
= 1 - f_L(x,\varepsilon) - f_T(x,\varepsilon)$.

The retarded matrix Green function in terms of the position and
energy dependent normal $g(\varepsilon,x)$ and anomalous
$F_i(\varepsilon,x)$ Green functions is:
\begin{equation}\label{eqn:Retarded}
    \hat{G}^R = \left( \begin{array}{cc} g(\varepsilon,x) & F_1(\varepsilon,x) \\
    F_2(\varepsilon,x) & -g(\varepsilon,x) \\
    \end{array} \right)
\end{equation}
\noindent Substituting this in the retarded part of the Usadel
equation: $\hbar D \nabla(\hat{G}^R \nabla \hat{G}^R) = -i
[\hat{H}, \hat{G}^R]$ and using the normalization condition ($g^2
+ F_1F_2 = 1$), we find the retarded Usadel equations:
\begin{equation}\label{eqn:RetUsadel}
    \begin{array}{ll}
      \hbar D [g\nabla^2 F_1 - F_1\nabla^2 g] &= -2i\Delta g - 2i\varepsilon F_1 \\
      \hbar D [F_1\nabla^2 F_2 - F_2\nabla^2 F_1] &= 2i\Delta F_2  + 2i\Delta^* F_1
    \end{array}
\end{equation}
\noindent The second equation is essential when calculating the
non-equilibrium properties of superconductors. Its left-hand-side
is proportional to the divergence of the spectral
(energy-dependent) supercurrent, which is (compared to the
equilibrium case) no longer a conserved quantity.
A general relation between the advanced matrix Green function and
the retarded matrix Green function is given by: $\hat{G}^R =
-\tau_3 (\hat{G}^A)^\dagger \tau_3$. Using this, the Keldysh
matrix Green function $\hat{G}^K$ can be written entirely in terms
of $g$, $F_1$, $F_2$, $f_L$ and $f_T$:
\begin{equation}\label{eqn:Keldysh}
    \hat{G^K} = \left( \begin{array}{cc} (g+g^\dagger)f_{+} & F_1 f_{-} - F_2^\dagger f_{+} \\
    F_2 f_{+} - F_1^\dagger f_{-} & -(g+g^\dagger)f_{-} \\
    \end{array} \right)\\
\end{equation}
\noindent where $f_{\pm}=f_L\pm f_T$. Working out the kinetic part
of the Usadel equation: $\hbar D \nabla(\hat{G}^R \nabla \hat{G}^K
+ \hat{G}^K \nabla \hat{G}^A) = -i [\hat{H}, \hat{G}^K]$ we find
(combining the diagonal components) the kinetic equations
describing the non-equilibrium part:
\begin{equation}\label{eqn:KineticUsadel}
    \begin{array}{l}
      \hbar D\nabla j_\mathrm{energy} = 0 \\
      \hbar D\nabla j_\mathrm{charge} = 2R_L f_L + 2R_T f_T \\
    \end{array}
\end{equation}
The various elements in Eq. \ref{eqn:KineticUsadel} are given by:
\begin{equation}\label{eqn:Coefficients}
    \begin{array}{ll}
      j_\mathrm{energy} &= \Pi_L\nabla f_L + \Pi_X\nabla f_T + j_\varepsilon f_T\\
      j_\mathrm{charge} &= \Pi_T\nabla f_T - \Pi_X\nabla f_L + j_\varepsilon f_L\\
      \vspace{-9pt}\\
      \Pi_L &= \frac{1}{4}(2 + 2\abs{g}^2 - \abs{F_1}^2 - \abs{F_2}^2 )\\
      \Pi_T &= \frac{1}{4}(2 + 2\abs{g}^2 + \abs{F_1}^2 + \abs{F_2}^2 )\\
      \Pi_X &= \frac{1}{4}(\abs{F_1}^2 - \abs{F_2}^2 )\\
      \vspace{-10pt}\\
      j_\varepsilon &= \frac{1}{2}\Re\{F_1\nabla F_2 - F_2\nabla F_1\}\\
      R_L &= -\frac{1}{2}\Im\{\Delta F_2 + \Delta F_1^\dagger\}\\
      R_T &= -\frac{1}{2}\Im\{\Delta F_2 - \Delta F_1^\dagger\}\\
    \end{array}
\end{equation}
\noindent Equations \ref{eqn:KineticUsadel} are two coupled
diffusion equations for $f_L$ and $f_T$, describing the
divergences in the spectral energy current and the spectral charge
current. The total charge current is given by
$J=\frac{1}{2e\rho}\int j_\mr{charge}d\varepsilon$ with $\rho$ the
resistivity. The terms $\Pi_L$ and $\Pi_T$ can be related to an
effective diffusion constant for the energy and charge mode
respectively and $\Pi_X$ as a "cross-diffusion" between them.
$j_\varepsilon$ is the spectral supercurrent and $R_L$ and $R_T$
describe the "leakage" of spectral current to different energies,
where the total leakage-current $\propto \int [R_L f_L + R_T f_T]
d\varepsilon$ is zero. In the small signal limit the terms
$\Pi_X$, $j_\varepsilon$ and $R_L$ are small and can in many cases
be neglected (linear approach), effectively decoupling $f_L$ and
$f_T$. In this article we go beyond this limit.

The Usadel equation is supplemented by a self-consistency
relation:
\begin{equation}\label{eqn:SelfCons}
    \hat{H}_{(1,2)} = \frac{N_0 V_\mathrm{eff}}{4}\int_{-\hbar \omega_D}^{\hbar \omega_D}\hat{G}_{(1,2)}^{K}d\varepsilon
\end{equation}
\noindent Here, $N_0$ is the normal density of states around the
Fermi energy, $V_\mathrm{eff}$ the effective attractive
interaction and the integral limits are set by the Debye energy
$\hbar \omega_D$. The resulting equation for $\Delta$ becomes:
$\Delta = -\frac{1}{4}N_0 V_\mathrm{eff}\int_{-\hbar
\omega_D}^{\hbar \omega_D}[(F_1 - F_2^\dagger)f_L - (F_1 +
F_2^\dagger)f_T ]d\varepsilon$.

To calculate spectral and transport properties, one needs to know
the self-consistent solution of $\Delta$. In most practical cases,
this has to be done numerically. A convenient solution scheme is
to first find the Green functions of the system by solving the
retarded equations for a certain $\Delta$, next to determine the
quasiparticle distribution functions by solving the kinetic
equations and then calculate a new $\Delta$ using the
self-consistency relation. This process has to be repeated until
$\Delta$ converges. As a starting value for $\Delta$ we use the
BCS form at zero temperature. To simplify the calculations a
parameterization is used that automatically fulfills the
normalization condition. It is convenient to take $g =
\cosh(\theta)$, $F_1 = \sinh(\theta)e^{i\chi}$ and $F_2 =
-\sinh(\theta)e^{-i\chi}$, where $\theta$ and $\chi$ are position
and energy dependent (complex) variables. At the interfaces
between the superconducting wire and the normal metallic
reservoirs we use the following boundary conditions: $\theta =
\nabla\chi = 0$ (retarded equation) and $f_{L,T} =
\frac{1}{2}(\tanh\frac{\varepsilon + eV}{2k_B T}\pm
\tanh\frac{\varepsilon - eV}{2k_B T})$ (kinetic equation), where
the latter are the usual reservoir distribution functions.

\begin{figure}[t]
  \includegraphics[width=8cm]{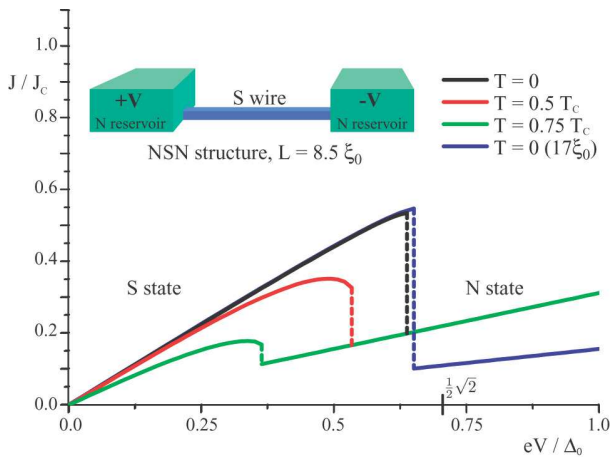}\\
  \caption{The calculated current(J)-voltage(V) relation of a
  superconducting wire of length $L=8.5\xi_0$ between normal metallic reservoirs (see inset) at several
  temperatures, and for a wire of length $17\xi_0$ at $T=0$. $J_c$ is the critical current density, and $\Delta_0$ the bulk gap energy}
  \label{fig:IVNSN}
\end{figure}
The transport properties of the NSN system can now be calculated
with the equations described above. In a previous analysis a
finite differential conductance was found at zero bias employing a
linear response calculation \cite{Boogaard2004}. With the approach
introduced here, the full current-voltage relation can be
obtained. The result at several temperatures is displayed in Fig.
\ref{fig:IVNSN}, with the voltage normalized to
$\Delta_0(=\Delta_\mr{bulk,T=0})$ and the current density
normalized to the critical current density $J_\mr{c}\approx
0.75\frac{\Delta_0}{\xi_0\rho e}$ \cite{Anthore2003}, with
$\xi_0=\sqrt\frac{\hbar D}{\Delta_0}$. At $\mathrm{T=0}$ we
observe a linear resistance at low voltages caused by the decay of
$f_T$ (Fig. \ref{fig:Distributions}c), and a critical point
(voltage) above which the resistance is equal to the normal state
resistance. At higher temperatures ($\mathrm{T=0.5, 0.75 T_c}$) a
linear approach would only give an adequate approximation in a
limited voltage range. We will argue below that the superconductor
switches to the normal state by $f_L$ which is controlled by the
voltage and cannot be interpreted as a critical current.

\begin{figure}[b]
  \includegraphics[width=8cm]{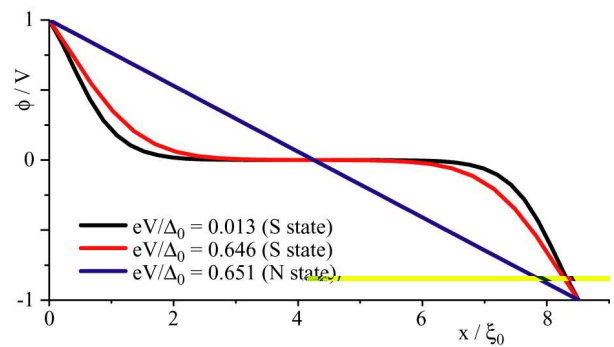}\\
  \caption{The normalized electrostatic potential $\phi$ as a function of
  position $x$ along the superconducting wire for bias voltages prior to and
  immediately after the transition (at $T=0$).}\label{fig:Vwire}
\end{figure}

In Fig. \ref{fig:Vwire} the electrostatic potential
$\phi=\int_0^\infty f_T\Re\{g\}d\epsilon$ along the wire is shown
at zero temperature prior to ($\mathrm{eV/\Delta_0}=0.013\mbox{,}
0.646$) and immediately after ($\mathrm{eV/\Delta_0=0.651}$) the
transition. The potential can be seen to drop to zero over a
distance of the order of the coherence length due to the normal-
to supercurrent conversion. This mechanism also gives rise to the
finite zero bias resistance. The profile hardly changes over the
full range of voltages, until the critical value is reached, after
which the electrostatic potential drops in a linear fashion,
indicating the system is in the normal state. The minimal changes
emphasize the limited influence of $f_T$ on the superconducting
state (i.e. on $\Delta$).

The current density at which the superconductor switches to the
normal state (for $\mathrm{T=0}$) is much smaller than the
critical current density in an infinitely long wire
($\mathrm{J}/\mathrm{J_c}=1$). Neither is the transition triggered
at the weaker superconducting edges as indicated by the shape of
the electrostatic potential profile in Fig. \ref{fig:Vwire}.

The parameter that determines whether or not the superconducting
state exist is $\mathrm{\Delta}$, as follows from Eq.
\ref{eqn:SelfCons}. The integral in this self-consistency equation
sums all pair-states (either occupied by a Cooper pair, or empty).
$F_i$ gives the Cooper pair density-of-states and $f_L$ and $f_T$
determine which of those states are doubly occupied or doubly
empty and which are singly occupied (broken) due to the presence
of quasiparticles.
In equilibrium at $\mathrm{T=0}$, a switch to the normal state can
only be caused by reaching a critical phase gradient, entering
$\Delta$ via $F_i$. In the presence of quasiparticles, $\Delta$
(and thus potentially the state of the system), is also influenced
by the distribution functions. It was noticed above that the
charge mode $f_T$ has a very limited influence on $\Delta$. The
effect of the energy mode $f_L$ is examined below.

\begin{figure}[t]
  \includegraphics[width=8cm]{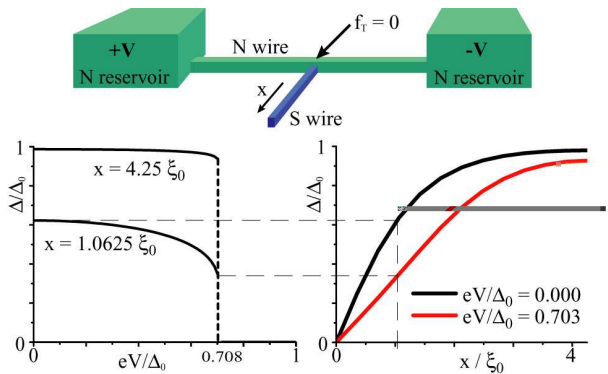}\\
  \caption{Top: T-shaped geometry, bottom: pair potential $\mathrm{\Delta}$ in the S-wire as
  function of (left) voltage at two different positions; (right) position at two different voltages.
  The breakdown voltage is at $eV/\Delta_0 = 0.707$}\label{fig:DeltaHshaped}
\end{figure}

By a small modification of our system to a T-shaped geometry as
shown in Fig. \ref{fig:DeltaHshaped}, we can in a direct way
disentangle the effects of $f_L$ and $f_T$ on $\mathrm{\Delta}$.
This setup can be thought of as the connection of the
superconducting wire to the center of a normal wire. In the middle
of such a wire $f_T$ is equal to zero, but $f_L$ is not. The
result for the pair potential at the edge of the superconducting
wire as a function of the voltage of the reservoirs is shown in
Fig. \ref{fig:DeltaHshaped}.
Although there is no net current flowing through the
superconductor, at a certain voltage the pair potential collapses.
The voltage that is necessary to trigger this transition to the
normal state is very close to the transition in Fig.
\ref{fig:IVNSN} (where we used the two terminal setup). Apparently
the influence of $f_L$ is important, since it can cause the
superconductor to switch to the normal state irrespective of the
value of the supercurrent. Clearly the influence of $f_L$ on the
state of the superconductor is larger than the influence of the
supercurrent on this same quantity.

The quantity that defines the possible states of the system is the
free energy. Evidently the superconductor compares two states for
the minimization of this free energy: the first state is the
superconducting state in which the free energy remains constant as
a function of voltage (and independent of the shape of $f_L$
provided this shape does not change for energies larger than the
gap). The second possible state is the normal state. At zero
temperature, in the absence of a bias voltage, the difference in
free energy between the two states is the condensation energy of
the superconductor. When the voltage is increased (but still
$\mathrm{eV} < \Delta$), the free energy of the superconducting
state remains constant while the free energy of the normal state
decreases since in that case electrons occupy higher energy states
due to the applied voltage.

To illustrate the effect, we calculate explicitly the free energy
difference between the superconducting state ($\mr{F_S}$) and
normal state ($\mr{F_N}$) at zero temperature for a bulk
superconductor (analytically) and the T-shaped structure
(numerically), as a function of voltage (which appears in $f_L$)
and $\Delta$. At zero temperature the free energy of the system
reduces to the internal energy (kinetic plus potential)
\cite{Kopnin2001}.
\begin{figure}[htb]
  \includegraphics[width=8cm]{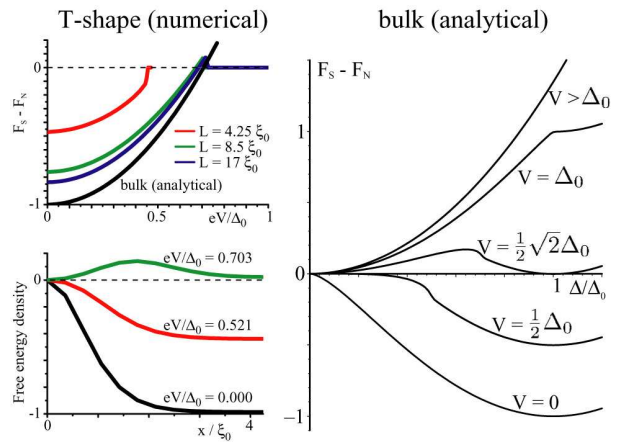}\\
  \caption{Free energy difference between the superconducting and normal state.
  Right: analytical bulk solution showing the bistable voltage range.
  Left: numerical solutions for (top) increasing wire length as function of voltage and
  (bottom) as function of position. Free energies are normalized to $\mathrm{H_c^2(0)/8\pi}$
  }\label{fig:Freeenergy}
\end{figure}
From the analytical calculation for the bulk following Bardeen
\cite{Bardeen1962} we find that $f_L$ changes the free energy in
such a way that at $\mathrm{eV} = \frac{1}{2}\sqrt{2}\Delta_0$ the
superconductor undergoes a first order phase transition. For the
voltage range $\frac{1}{2}\Delta_0 < \mathrm{eV} < \Delta_0$ the
state of the system has two solutions. The free energy difference
for the bulk superconductor is shown in Fig. \ref{fig:Freeenergy}
\footnote{Hysteretic behavior due to the first order transition is
also present in the numerical calculation, for clarity in Fig.
\ref{fig:IVNSN} and \ref{fig:DeltaHshaped} only the upsweeps are
displayed.}. Numerical results for the free energy of the T-shaped
geometry are shown as well (both as function of position and as
function of voltage). For long wires, the numerical results
approach the analytical (bulk) calculation. This indicates that
the effect of the bias voltage can indeed be related to the
existence of a first order phase transition at zero temperature.

In conclusion, we have studied the role of the energy mode $f_L$
of the quasiparticle distribution on the properties of a
superconducting nanowire. We employ a numerical simulation of the
Usadel equation in full-response and find a non-thermal
distribution for $f_L$ (caused by an applied bias voltage) which
drives a first order transition from the superconducting state to
the normal state irrespective of the current. A direct calculation
on the free energy of a bulk superconductor confirms that the
voltage indeed causes the phase transition. In general, the
significant role played by $f_L$ found in these superconducting
nanowires stresses the importance of treating $f_L$ and $f_T$ on
equal footing.

This work is part of the research programme of the 'Stichting voor
Fundamenteel Onderzoek der Materie (FOM)', which is financially
supported by the 'Nederlandse Organisatie voor Wetenschappelijk
Onderzoek (NWO)'. We thank W. van Saarloos for discussion.



\begin{thebibliography}{8}
\expandafter\ifx\csname
natexlab\endcsname\relax\def\natexlab#1{#1}\fi
\expandafter\ifx\csname bibnamefont\endcsname\relax
  \def\bibnamefont#1{#1}\fi
\expandafter\ifx\csname bibfnamefont\endcsname\relax
  \def\bibfnamefont#1{#1}\fi
\expandafter\ifx\csname citenamefont\endcsname\relax
  \def\citenamefont#1{#1}\fi
\expandafter\ifx\csname url\endcsname\relax
  \def\url#1{\texttt{#1}}\fi
\expandafter\ifx\csname
urlprefix\endcsname\relax\def\urlprefix{URL }\fi
\providecommand{\bibinfo}[2]{#2}
\providecommand{\eprint}[2][]{\url{#2}}

\bibitem[{\citenamefont{Pothier et~al.}(1997)\citenamefont{Pothier, Gu\'{e}ron,
  Birge, Esteve, and Devoret}}]{Pothier1997}
\bibinfo{author}{\bibfnamefont{H.}~\bibnamefont{Pothier}},
  \bibinfo{author}{\bibfnamefont{S.}~\bibnamefont{Gu\'{e}ron}},
  \bibinfo{author}{\bibfnamefont{N.~O.} \bibnamefont{Birge}},
  \bibinfo{author}{\bibfnamefont{D.}~\bibnamefont{Esteve}}, \bibnamefont{and}
  \bibinfo{author}{\bibfnamefont{M.~H.} \bibnamefont{Devoret}},
  \bibinfo{journal}{Phys.\ Rev.\ Lett.} \textbf{\bibinfo{volume}{79}},
  \bibinfo{pages}{3490} (\bibinfo{year}{1997}).

\bibitem[{\citenamefont{Pierre et~al.}(2003)\citenamefont{Pierre, Gougam,
  Anthore, Pothier, Esteve, and Birge}}]{Birge2003}
\bibinfo{author}{\bibfnamefont{F.}~\bibnamefont{Pierre}},
  \bibinfo{author}{\bibfnamefont{A.~B.} \bibnamefont{Gougam}},
  \bibinfo{author}{\bibfnamefont{A.}~\bibnamefont{Anthore}},
  \bibinfo{author}{\bibfnamefont{H.}~\bibnamefont{Pothier}},
  \bibinfo{author}{\bibfnamefont{D.}~\bibnamefont{Esteve}}, \bibnamefont{and}
  \bibinfo{author}{\bibfnamefont{N.~O.} \bibnamefont{Birge}},
  \bibinfo{journal}{Phys.\ Rev.\ B} \textbf{\bibinfo{volume}{68}},
  \bibinfo{pages}{085413} (\bibinfo{year}{2003}).

\bibitem[{\citenamefont{\mbox{For a review, see}
  Nikolai~Kopnin}(2001)}]{Kopnin2001}
\bibinfo{author}{\bibnamefont{\mbox{For a review, see} Nikolai~Kopnin}},
  \emph{\bibinfo{title}{Theory of nonequilibrium superconductivity}}, vol.
  \bibinfo{volume}{110} of \emph{\bibinfo{series}{International series of
  monographs on physics}} (\bibinfo{publisher}{Oxford University Press},
  \bibinfo{year}{2001}), \bibinfo{note}{and references therein}.

\bibitem[{Gau()}]{Gauge}
\bibinfo{note}{Where we introduce the phase $\chi$: $\phi = \chi -
  \frac{2e}{\hbar}\int_0 {x A(l) dl}$}.

\bibitem[{\citenamefont{Schmid and Sch\mbox{\"{o}}n}(1975)}]{Schmid197A}
\bibinfo{author}{\bibfnamefont{A.}~\bibnamefont{Schmid}} \bibnamefont{and}
  \bibinfo{author}{\bibfnamefont{G.}~\bibnamefont{Sch\mbox{\"{o}}n}},
  \bibinfo{journal}{J.\ Low Temp.\ Phys.} \textbf{\bibinfo{volume}{20}},
  \bibinfo{pages}{207} (\bibinfo{year}{1975}).

\bibitem[{\citenamefont{Boogaard et~al.}(2004)\citenamefont{Boogaard,
  Verbruggen, Belzig, and Klapwijk}}]{Boogaard2004}
\bibinfo{author}{\bibfnamefont{G.~R.} \bibnamefont{Boogaard}},
  \bibinfo{author}{\bibfnamefont{A.~H.} \bibnamefont{Verbruggen}},
  \bibinfo{author}{\bibfnamefont{W.}~\bibnamefont{Belzig}}, \bibnamefont{and}
  \bibinfo{author}{\bibfnamefont{T.~M.} \bibnamefont{Klapwijk}},
  \bibinfo{journal}{Phys.\ Rev.\ B} \textbf{\bibinfo{volume}{69}},
  \bibinfo{pages}{220503} (\bibinfo{year}{2004}).

\bibitem[{\citenamefont{Anthore et~al.}(2003)\citenamefont{Anthore, Pothier,
  and Esteve}}]{Anthore2003}
\bibinfo{author}{\bibfnamefont{A.}~\bibnamefont{Anthore}},
  \bibinfo{author}{\bibfnamefont{H.}~\bibnamefont{Pothier}}, \bibnamefont{and}
  \bibinfo{author}{\bibfnamefont{D.}~\bibnamefont{Esteve}},
  \bibinfo{journal}{Phys.\ Rev.\ Lett.} \textbf{\bibinfo{volume}{90}},
  \bibinfo{pages}{127001} (\bibinfo{year}{2003}).

\bibitem[{\citenamefont{Bardeen}(1962)}]{Bardeen1962}
\bibinfo{author}{\bibfnamefont{J.}~\bibnamefont{Bardeen}},
  \bibinfo{journal}{Rev.\ Mod.\ Phys.} \textbf{\bibinfo{volume}{34}},
  \bibinfo{pages}{667} (\bibinfo{year}{1962}).

\end{thebibliography}

\end{document}